\documentclass[twocolumn,prb,showpacs]{revtex4}

\usepackage{graphicx}
\usepackage{amsmath,amssymb,amsfonts}

\newcommand{\ket}[1]{\lvert #1\rangle}
\newcommand{\bra}[1]{\langle#1 \rvert}
\newcommand{\abs}[1]{\lvert #1 \rvert}

\newcommand{\braket}[2]{\langle #1 \rvert #2\rangle}

\begin{document}
\title{Benchmarking GW against exact diagonalization for semi-empirical models}


\author{K. Kaasbjerg$^1$ and K. S. Thygesen$^2$}
\affiliation{$^1$Nano-Science Center, Niels Bohr Institute, University of Copenhagen \\
Universitetsparken 5, DK-2100 Copenhagen, Denmark \\
$^2$Center for Atomic-scale Materials Design (CAMD), \\
Department of Physics, Technical University of Denmark, DK - 2800 Kgs. Lyngby, Denmark}
\date{\today}

\begin{abstract}
  We calculate groundstate total energies and single-particle excitation
  energies of seven pi conjugated molecules described with the semi-empirical
  Pariser-Parr-Pople (PPP) model using self-consistent many-body perturbation
  theory at the GW level and exact diagonalization. For the total energies GW
  captures around 65$\%$ of the groundstate correlation energy. The lowest lying
  excitations are overscreened by GW leading to an underestimation of electron
  affinities and ionization potentials by approximately 0.15 eV corresponding to
  2.5$\%$. One-shot G$_0$W$_0$ calculations starting from Hartree-Fock reduce
  the screening and improve the low-lying excitation energies. The effect of the
  GW self-energy on the molecular excitation energies is shown to be similar to
  the inclusion of final state relaxations in Hartree-Fock theory. We discuss
  the break down of the GW approximation in systems with short range
  interactions (Hubbard models) where correlation effects dominate over
  screening/relaxation effects. Finally we illustrate the important role of the
  derivative discontinuity of the true exchange-correlation functional by
  computing the exact Kohn-Sham levels of benzene.
\end{abstract}

\maketitle 

\section{Introduction}
For more than two decades the many-body GW approximation of Hedin\cite{hedin}
has been the state of the art for calculating band structures of metals,
semiconductors, and insulators\cite{hybertsen86,wilkins,gunnarsson,orr}. With
the entry of nanoscience the use of the GW method has been extended to low-dimensional
systems such as molecules, carbon nanotubes, graphene and molecule-surface
interfaces~\cite{grossman01,stan_eulett,louie,reining,Chelikowsky,neaton,Freysoldt}.
In these systems the interplay between quantum confinement (in one or more
dimensions) and electronic correlation effects leads to novel phenomena like the
renormalization of molecular electronic levels at surfaces by dynamical
polarization in the substrate\cite{neaton,kaasbjerg,thygesen09,Freysoldt}. Very
recently, the non-equilibrium version of the GW approximation has been applied
to quantum transport and dynamics in molecular
junctions\cite{spataru,thygesen_jcp,darancet,thygesen_prl,leeuwen_trans,spataru2,verdozzi,leeuwen_dynamics}
where dynamic correlations seems to be particularly important.

As the range of systems to which the GW approximation is being applied continues
to expand, critical investigations of the performance of GW for other systems
than the crystalline solids become important. Here we report on benchmark GW
calculations for $\pi$-conjugated molecules based on the semi-empirical
Pariser-Parr-Pople (PPP) model~\cite{Pople,PariserParrI,PariserParrII}. By
comparing with exact results we obtain a direct and unbiased estimate of the
quality of the GW approximation in molecular systems.

Previous benchmark model studies of the GW approximation have all focused on
Hubbard models with local
interactions~\cite{schindlmayr1,schindlmayr2,verdozzi_prl,verdozzi} with the
conclusion that GW works well for small interaction strengths but fails for
larger interactions strength. The use of GW in systems with local interactions
is in fact unfortunate because the importance of electronic screening, which is
the main effect described by GW, is weak in comparison to correlation effects.
In contrast to Hubbard models, the PPP description includes long range
interactions and its parameters have been fitted to yield realistic excitation
energies of conjugated molecules. It therefore provides a better and more
natural starting point for a study addressing the accuracy of GW for real
molecules and nanostructures. We mention that in a related work we have
performed first-principles GW calculations for a series of 33 molecules arriving
at very similar conclusions regarding the performance of GW as those reported
here.\cite{paperII}

\emph{Ab-initio} GW calculations typically involve a number of "technical"
approximations such as the plasmon pole approximation, the neglect of
off-diagonal matrix elements in the GW self-energy, or analytic continuations.
Moreover they are usually performed non-selfconsistently and are subject to
basis set errors. In the present work the GW calculations are carried out fully
self-consistently without any further approximations apart from the GW
approximation itself. 

In this work we calculate total energies and excitation spectra of the seven
conjugated molecules listed in Tab.~\ref{tab:Entropies}. The excitation spectrum
of a system can be obtained from the spectral function
\begin{align}
  \label{eq:SpectralFunction}  
  A_i(\varepsilon) = 2\pi \sum_n \bigg[ &
      \abs{\bra{\Psi_n(N+1)}c_i^{\dagger}\ket{\Psi_0(N)}}^2
      \delta(\varepsilon-\varepsilon_n)  \nonumber \\
    & +\abs{\bra{\Psi_n(N-1)}c_i^{\phantom\dagger}\ket{\Psi_0(N)}}^2 
      \delta(\varepsilon-\varepsilon_n) \bigg] ,
\end{align}
which has peaks at the excitation energies $\varepsilon_n = E_n(N+1)-E_0(N)$ and
$\varepsilon_n = E_0(N)-E_n(N-1)$ corresponding to electronic addition and
removal energies, respectively. Often in the GW literature, excitation energies
are often referred to as quasi-particle (QP) energies. In the expressions for
the excitation energies $E_n(N)$ denotes the energy of the $n$th excited
$N$-electron state, $\ket{\Psi_n(N)}$, with $N$ referring to the neutral state
of the system. For molecules the first addition and the first removal energy,
i.e. $n=0$, corresponds to the electron affinity and the ionization potential.
In Hartree-Fock theory Koopman's theorem~\cite{Fulde} states that the
eigenvalues of the Hartree-Fock Hamiltonian equal the addition/removal energies
calculated without orbital relaxations in the charged states, i.e.
$\varepsilon_n^{\text{HF}}=\langle
c^\dagger_n\Psi_0^{\text{HF}}(N)|H|c^\dagger_n
\Psi_0^{\text{HF}}(N)\rangle-E_0^{\text{HF}}(N)$ for a virtual orbital $n$. In
particular, the highest occupied molecular orbital (HOMO) and the lowest
unoccupied orbital (LUMO) represent well defined approximations to the
ionization potential and electron affinities, respectively~\footnote{In the
  remaining of the paper HOMO and LUMO abbreviations will be used to refer to
  the ionization potential and electron affinity also outside Hartree-Fock
  theory.}. This approximation neglects two important effects. One is the
relaxation of the single-particle HF orbitals when an electron is removed from
or added to the molecule. The other is the correlation energy which by
definition is omitted in HF theory. It is instructive to write the exact QP
energies as the sum of the three contributions
\begin{equation}
  \label{eq:QuasiLevels}
  \varepsilon_n = \varepsilon_n^{\text{HF}} + \Delta_{\text{relax}} +
                  \Delta_{\text{corr}} ,
\end{equation}
The relaxation contribution is the correction that follows by calculating the
QP energy from self-consistently determined HF energies of the
neutral \emph{and} the charged states $N\pm 1$. The last term
$\Delta_{\text{corr}}$ is the remaining contribution from the correlation
energy. For the addition of an electron, i.e. an unoccupied orbital, the
relaxation and correlation contributions are given by
\begin{equation}
  \label{eq:Delta_relax}
  \Delta_{\text{relax}} = E_n^{\text{HF}}(N+1) - E_0^{\text{HF}}(N) -
\varepsilon_n^{\text{HF}} 
\end{equation}
and 
\begin{equation}
  \label{eq:Delta_corr}
  \Delta_{\text{corr}} = [E_{n}(N+1)-E_n^{\text{HF}}(N+1)] - [E_{0}(N)-
                         E_0^{\text{HF}}(N)] .
\end{equation}

In extended systems the potential due to a single delocalized electron/hole
decreases with the size of the system. Hence, in such systems there will be no
or little relaxation of the states due to the addition/removal of an electron,
and the majority of the correction to the QP energy will come from the
correlation part $\Delta_{\text{corr}}$. In molecules, nanostructures, molecules
at surfaces, and disordered systems with finite localization lengths, this is
not the case. Here, the introduction of an additional electron or hole will lead
to a relaxation of the single-particle orbitals corresponding to a screening of
the additional charge. As a consequence, the relaxation correction
$\Delta_{\text{relax}}$ to the QP energy cannot be neglected in such systems. In
fact, we find that $\Delta_{\text{relax}}$ is larger than $\Delta_{\text{corr}}$
for all the molecules studied here, and that the GW excitation energies
correspond roughly to including only $\Delta_{\text{relax}}$ in Eq.
(\ref{eq:QuasiLevels}).

The paper is outlined as follows. In Sec. \ref{sec:ppp} the PPP model
Hamiltonian for conjugated molecules is introduced. In Secs.
\ref{sec:method_gw} and \ref{sec:method_ex} we provide an overview of
the theory and numerical implementation of the GW and exact
calculations, and in Sec.  \ref{sec:neuman} we discuss the use of the
von Neumann entropy as a measure of correlation. The results for total
energies and spectral properties of the PPP model are presented in
Secs. \ref{sec:total} and \ref{sec:spec}, and a comparison is made to
short ranged Hubbard models in Sec. \ref{sec:hubbard}. In
Sec.\ref{sec:latticedft} we calculate the exact Kohn-Sham levels for the benzene molecule and
compare to the exact QP levels. The
conclusions are given in Sec. \ref{sec:conclusions}.

\section{Pariser-Parr-Pople Hamiltonian}\label{sec:ppp}
The Pariser-Parr-Pople model is an effective $\pi$-electron description of
conjugated molecules that includes electron-electron interactions explicitly.
The PPP Hamiltonian is given by
\begin{align}
  \label{eq:PPPHamiltonian}
  H = & \sum_i \varepsilon_i \hat{n}_i
        - \sum_{\langle ij\rangle\sigma}t_{ij} 
        c^{\dagger}_{i\sigma}c^{\phantom\dagger}_{j\sigma}
        \nonumber\\
      & + \frac{1}{2} \sum_{i\neq j} V_{ij} (\hat{n}_i-Z_i)(\hat{n}_j-Z_j)  
        + \sum_i U_i \hat{n}_{i\uparrow}\hat{n}_{i\downarrow} ,
\end{align}
where $c^{\dagger}_i$ ($c_i$) creates (annihilates) an
electron in the $p_z$ orbital on atom $i$ of the molecule, $\hat{n}_i =
\hat{n}_{i\uparrow} + \hat{n}_{i\downarrow}$ is the number operator,
$\hat{n}_{i\sigma} = c^{\dagger}_{i\sigma} c^{\phantom\dagger}_{i\sigma}$, $Z_i$
is the valence (i.e. the number of $\pi$ electrons) of atom $i$, and $\langle
ij \rangle$ denotes nearest neighbour hopping. The Ohno
parametrization~\cite{Ohno} is used for the long range interactions
\begin{equation}
  \label{eq:V_ij}
  V_{ij} =  \frac{14.397}{\sqrt{(28.794/(U_i + U_j))^2 + R_{ij}^2}} \; ,
\end{equation}
where $R_{ij}$ is the inter-atomic distance (in \AA) and $U_i$ is the onsite
Coulomb interaction (in eV). For large distances the Ohno parametrization
recovers the $1/r$ behavior of the Coulomb interaction while it for small
distances represents a screened interaction that interpolates to onsite Coulomb
interaction $U_i$ for $R_{ij} = 0$. The onsite energy $\varepsilon_i$, the
hopping element $t_{ij}$ and the onsite Coulomb interaction $U_i$ are treated as
fitting parameters. In the present work values for these parameters have been
taken from the
literature~\cite{soos-transferintegral,Barford-PPPandPPV,Barford-Benzene,Barford-PPP,Barford-PPV}.
Since existing parameters have been optimized to optical excitation spectra, an
exact agreement with experimental values for the molecular gaps is not to be
expected.

\section{Methods}

\subsection{GW approximation}\label{sec:method_gw}
Hedin's equations~\cite{hedin} provides a formally exact framework for the
determination of the single-particle Green function in a self-consistent manner.
In the GW approximation, which follows by neglecting the so called vertex
corrections, the electronic self-energy $\Sigma$ is given by the product of the
Green function $G$ and the screened interaction $W$, and can be written
symbolically as
\begin{equation}
  \label{eq:SelfEnergy}
  \Sigma = i G W,
\end{equation}
where the Green function obeys the usual Dyson equation $G=G_0 + G_0\Sigma G$.
The screened interaction $W$ is given by the bare Coulomb interaction $V$ and
the polarization in the random-phase approximation (RPA) $P=-iGG$ through the
Dyson-like equation
\begin{equation}
  \label{eq:W}
  W = V + VPW.
\end{equation}
In fully self-consistent GW the set of coupled equations for $\Sigma$, $G$, $P$,
and $W$ are solved iteratively until the Green function has converged. Due to
the computational requirement of a fully self-consistent GW scheme,
\textit{ab-initio} GW calculations are usually carried out non-selfconsistently.
This approach, which is referred to as G$_0$W$_0$, starts from an approximate
$G_0$, typically the non-interacting Kohn-Sham Green function, from which a
single self-energy iteration is carried out to obtain the final Green function.

\subsubsection{Numerical details}
The GW calculations have been performed following the method described in detail
in Ref. \onlinecite{thygesen_gw_prb}. Here we give a brief overview of the method
for completeness.

The retarded and advanced single-particle Green functions are given by
\begin{equation}
G^{r/a}(\varepsilon)=(\varepsilon\pm
i\eta-H_0-V_{\text{H}}-\Sigma_{GW}(\varepsilon))^{-1}
\end{equation}
where $\eta$ is a small positive infinitesimal, $H_0$ contains the first two
terms in Eq. \eqref{eq:PPPHamiltonian}, and $V_{\text{H}}$ is the Hartree
potential. We represent the Green functions and all other energy-dependent
quantities on a uniform grid, $-E_m,-E_m+d\varepsilon,\ldots,E_m$. The Fast
Fourier Transform is used to switch between the energy and time representations.
Since $\eta$ determines the minimum width of features in the Green function's
energy dependence, the energy grid spacing should obey $d\varepsilon \ll \eta$.
All results presented here have been converged with respect to
$\eta,d\varepsilon,E_m$. Typical converged values are (in eV)
$\eta=0.02,d\varepsilon=0.005,E_m=50$.

The lesser/greater Green functions are given by
\begin{eqnarray}
G^<(\varepsilon)&=&-f(\varepsilon-\mu)[G^r-G^a]\\
G^>(\varepsilon)&=&(1-f(\varepsilon-\mu))[G^r-G^a]
\end{eqnarray}
where $f(\varepsilon-\mu)$ is the Fermi-Dirac function. The chemical potential
$\mu$ is adjusted to yield the desired number of electrons in the system. The
formulation in terms of a fixed chemical potential rather than a fixed particle
number is reminiscent of the fact that the method has been developed for quantum
transport. The one-body density matrix is given by
\begin{equation}
\rho_{ij} = -i\int G^<_{ij}(\varepsilon)d\varepsilon.
\end{equation}
From $\rho$ the Hartree and exchange potentials follow 
\begin{eqnarray}
V_{\text{H},ij} & = & \delta_{ij} 2\sum_k V_{ik}\rho_{kk} \\
V_{x,ij}        & = & -V_{ij}\rho_{ij},
\end{eqnarray}
where we have defined $V_{ii}=U_i$, see Eq. (\ref{eq:PPPHamiltonian}).
 
The retarded/advanced and lesser/greater components of the quantities needed to
construct the GW self-energy read\cite{thygesen_gw_prb}
\begin{eqnarray}
\Sigma^{</>}_{\text{GW},ij}(t)&=&iG^{</>}_{ij}(t)W^{</>}_{ij}(t)\\
W_{ij}^{</>}(\varepsilon)&=&\sum_{kl}W_{ik}^r(\varepsilon)P_{kl}^{</>}(\varepsilon)W_{lj}^a(\varepsilon)\\
\label{eq:Wr}
W^{r/a}_{ij}(\varepsilon)&=&\sum_{k}P^{r/a}_{ik}[1-V P^{r/a}(\varepsilon)]^{-1}_{kj}\\
P_{ij}^{</>}(t)&=&G_{ij}^{</>}(t)G_{ji}^{>/<}(-t)
\end{eqnarray}
The GW equations have been expressed in the time or energy domain according to
where they are simplest. This also reflects the practical implementation.

The retarded components of $\Sigma_{\text{GW}}$ and $P$ are obtained using the
fundamental relation
\begin{equation}
F^r(t)=-i\theta(t)[F^>(t)-F^<(t)]
\end{equation}
which is the Kramers-Kronig relation in the time domain relating the imaginary
and real parts of $F^r$.

Since the GW self-energy depends on the Green function and vice versa, the
equations must be iterated until self-consistency. To speed up convergence we
use the Pulay mixing scheme\cite{pulay} as described in
Ref.~\onlinecite{thygesen_gw_prb}.

\subsubsection{Total energy}

The total energy can be split into kinetic (and external), Hartree, and
exchange-correlation energy $E=E_0+E_{\text{H}}+E_{\text{xc}}$. In terms of the
Green function we have
\begin{equation}
  \label{eq.E0}
  E_0+E_{\text{H}} = \text{Tr}[H_0 \rho]+\frac{1}{2}\text{Tr}[V_{\text{H}} \rho]
\end{equation}
For the exchange-correlation energy we have 
\begin{equation}
  \label{eq.xc}
  E_{\text{xc}} = \frac{1}{2i}\int 
  \text{Tr}[\Sigma^r(\varepsilon)G^<(\varepsilon)+\Sigma^<(\varepsilon)G^a(\varepsilon)]
d\varepsilon,
\end{equation}
where $\Sigma$ is the exchange-correlation self-energy. In this work $\Sigma$ is
either the bare exchange, $\Sigma_x$, yielding the HF approximation, or the GW
self-energy, $\Sigma_{\text{GW}}$. The expression (\ref{eq.xc}) follows by
expressing $\langle \hat V \rangle$ in terms of the two-particle Green function,
$G_2$, and then using the defining equation for the self-energy in terms of
$G_2$\cite{fetter_walecka}.

\subsection{Exact diagonalization}\label{sec:method_ex}
The most direct way to the spectral properties of a system is via the Lehmann
representation of the Green function in Eq. \eqref{eq:SpectralFunction}. However,
since this requires the full set of eigenstates and eigenvalues of the
Hamiltonian, it is of limited practical use and other routes must be taken. The
following section gives a brief overview of the Lanczos method for iterative
diagonalization of large matrices.

\subsubsection{Calculating the ground state - Lanczos algorithm}
In exact diagonalization the given many-body Hamiltonian is diagonalized
directly in the Fock space which is spanned by many-particle states (Slater
determinants). Since the dimensionality of the Fock space grows exponentially
with the number of basis orbitals, symmetries of the Hamiltonian can help to
reduce the dimensionality considerably. For the Pariser-Parr-Pople Hamiltonian
in Eq.~\eqref{eq:PPPHamiltonian} the number of up and down electrons,
$N_{\uparrow}$ and $N_{\downarrow}$, are good quantum numbers since their
corresponding operators commute with the Hamiltonian. This implies that the
exact diagonalization can be carried out in each of the
$(N_{\uparrow},N_{\downarrow})$-subblocks of the Fock space independently. The
dimensionality of each $(N_{\uparrow},N_{\downarrow})$-subblock is given by the
number of ways $N_{\uparrow}$ spin up electrons and $N_{\downarrow}$ spin down
electrons can be distributed over $L$ basis orbitals,
\begin{equation}
  \label{eq:HilbertDimension}
  d(N_{\uparrow},N_{\downarrow}) = \frac{L!}{N_{\uparrow}!(L-N_{\uparrow})!} \times
                                \frac{L!}{N_{\downarrow}!(L-N_{\downarrow})!}.
\end{equation}
Very often the ground state is located in the half-filled subblock, i.e.
$N_{\uparrow}=N_{\downarrow}=L/2$ where $L$ is the number of basis orbitals. For
$L=16$ the dimensionality of this subblock is $d=165636900$, implying that
storing a vector in double floating point precision requires $\sim
1\;\mathrm{Gb}$ of memory. With such memory requirements a full diagonalization
of the Hamiltonian is of course out of reach. If only the ground state is
needed, iterative methods can be employed. The basic idea of iterative methods
is to project the Hamiltonian onto the Krylov subspace $\mathcal{K}$ generated
by repeated applications of $H$ on an arbitrary initial state $\ket{\phi_0}$,
i.e.
\begin{equation}
  \label{eq:Krylov}
  \mathcal{K} = \mathrm{span} \{ \ket{\phi_0},H\ket{\phi_0},H^2\ket{\phi_0},
                                  \cdots,H^N\ket{\phi_0}
                              \} .
\end{equation}
In the Krylov subspace the extreme eigenvalues of the Hamiltonian converge fast
with respect to the size $N$ of the subspace, thus reducing the full
diagonalization to a manageable diagonalization of a $N\times N$ matrix, with
$N\ll d$.

In the Lanczos algorithm~\footnote{The present work has used the implementation
  from the IETL project~\cite{www:ietl}.} the Hamiltonian is projected onto a
specially constructed orthogonalised Krylov basis in which the Hamiltonian has a
tridiagonal representation. The basis vectors are generated recursively as
\begin{equation}
  \label{eq:LanczosBasis}
  \ket{\phi_{n+1}} = H\ket{\phi_n} - a_n \ket{\phi_n} - b_n^2\ket{\phi_{n-1}} ,
\end{equation}
where the coefficient are given by
\begin{equation}
  \label{eq:LanczosCoefficients}
  a_n = \frac{\bra{\phi_n}H\ket{\phi_n}}{\braket{\phi_n}{\phi_n}} 
  \quad \text{and} \quad 
  b_n^2 = \frac{\braket{\phi_n}{\phi_n}}{\braket{\phi_{n-1}}{\phi_{n-1}}}
\end{equation}
with initial conditions $b_0 = 0$ and $\ket{\phi_{-1}}=0$. At any point during
the Lanczos iterations only three Lanczos vectors needs to be kept in memory,
which makes the algorithm memory efficient. In the basis of the normalized
vectors (the basis vectors above are not normalized) the Hamiltonian has the
following tridiagonal representation
\begin{equation}
  \label{eq:TridiagonalHamiltonian}
  H =   
  \begin{pmatrix}
    a_0    & b_1    &  0     & \cdots & 0      \\
    b_1    & a_1    & b_2    &        & \vdots \\
    0      & b_2    & a_2    & \ddots & 0      \\
    \vdots &        & \ddots & \ddots & b_N    \\
    0      & \cdots & 0      & b_N    & a_N    \\
  \end{pmatrix}
\end{equation}
which can be readily diagonalized with methods for tridiagonal matrices. In
practice the Lanczos iterations are continued until the desired eigenvalues have
converged to a given tolerance. For the ground state energy $E_0$, typical
values for $N$ range from a few to $\sim 200$ depending on the system size.

The ground state resulting from a diagonalization of the tridiagonal Hamiltonian
in Eq.~\eqref{eq:TridiagonalHamiltonian} is provided in the Lanczos basis, i.e.
$\ket{\Psi_0}=\sum_n c_n\ket{\phi_n}$. In order to be able to calculate the
Green function, its representation in the original many-body basis is
required. Since the Lanczos vectors are not stored, the Lanczos iterations must
be repeated (starting from the same initial vector) to obtain the expansion
coefficients $\alpha_i=\sum_n c_n \braket{\Phi_i}{\phi_n}$ in the original
many-body basis $\{\ket{\Phi_i}\}_{i=1}^d$. 

The most time consuming part of the Lanczos algorithm is the matrix-vector
multiplication $H\ket{\phi_n}$. An efficient implementation of this part is
hence crucial. For this purpose it is convenient to use the bit representation
of an unsigned integer to code the basis states. Denoting the integers with bit
representations corresponding to the spin up and spin down occupations of a
given basis state with $I_{\uparrow}$ and $I_{\downarrow}$, respectively, the
integer representation of the basis state is $I = I_{\uparrow} + 2^L
I_{\downarrow}$. With the binary representation of the basis states, the
multiplication of the Hamiltonian can be done efficiently using bitwise
operations.

\subsubsection{Calculating the Green function}
Having obtained the ground state, the Green function can now be calculated. From
the Lehmann representation it follows that it can be written as
\begin{equation}
  G_{ij}^r(\varepsilon) = G_{ij}^e(\varepsilon) + G_{ij}^h(\varepsilon)
\end{equation}
with the electron and hole Green functions defined by
\begin{equation}
  \label{eq:ElectronGreensFunction}
  G_{ij}^e(\varepsilon) = \bra{\Psi_0^N}c_i^{\phantom\dagger} 
                     \frac{1}{\varepsilon-H+E_0^N+i\eta} 
                     c_j^{\dagger}\ket{\Psi_0^N}
\end{equation}
and
\begin{equation}
  \label{eq:HoleGreensFunction}
  G_{ij}^h(\varepsilon) = \bra{\Psi_0^N}c_j^{\dagger} 
                     \frac{1}{\varepsilon+H-E_0^N+i\eta} 
                     c_i^{\phantom\dagger}\ket{\Psi_0^N},
\end{equation}
respectively. In the following we focus on the electron Green function which is
the matrix representation of the resolvent operator $(z-H)^{-1}$ in the basis
spanned by the $\ket{i}=c_i^{\dagger}\ket{\Psi_0^N}$ vectors. To obtain the
$i$'th diagonal element,
\begin{equation}
  \label{eq:GFResolvent}
  G_{ii}^e(\varepsilon) = \bra{i}(z-H)^{-1}\ket{i}, 
\end{equation}
where $z=\varepsilon+E_0^N+i\eta$, again the Lanczos algorithm is used to put
$H$ on a tridiagonal form, but this time the Lanczos iterations are started from
the normalized initial state $\ket{\phi_0}=\ket{i}/b_0$ where
$b_0^2=\braket{i}{i}$. Hence, in the generated Krylov subspace the diagonal
element in Eq.~\eqref{eq:GFResolvent} corresponds to the matrix element
$b_0^2[(\varepsilon-H+E_0^N+i\eta)^{-1}]_{11}$ of a tridiagonal matrix, which
can be obtained as the continued fraction~\cite{Dagotto}
\begin{equation}
  \label{eq:ContinuedFraction}
  G_{ii}^e(\varepsilon) = \cfrac{b_0^2}{\varepsilon -a_0 - \cfrac{b_1^2}{\varepsilon
-a_1 -
      \cfrac{b_2^2}{\varepsilon -a_2 - \cdots}}} .
\end{equation}
Again the Lanczos iterations are continued until the frequency dependent Green
function element has converged.

\subsection{Von Neumann entropy}\label{sec:neuman}
The following section demonstrates how a quantitative measure of the degree of
correlations in a system can be obtained by considering the von Neumann entropy
of the reduced single-particle density matrix $\rho$. The entropy is defined by
\begin{equation}
  \label{eq:VonNeumannEntropy}
  S[\rho] = - \mathrm{Tr}[\rho \log{\rho}] = -\sum_n \rho_n \log{\rho_n},
\end{equation}
where in the last equality $\rho$ has been expressed in its diagonal
representation, $\rho =\sum_n \rho_n \ket{n}\bra{n}$. 

In the basis of the atomic $p_z$ orbitals the matrix elements of the
reduced density matrix are given by (with the spin index suppressed)
\begin{equation}
  \label{eq:ReducedDensityMatrix}
  \rho_{ij} = \bra{\Psi_0} c^{\dagger}_{j} c^{\phantom\dagger}_{i} 
              \ket{\Psi_0} ,
\end{equation}
with the diagonal elements equal to the site occupations. In the diagonal
representation $\rho_n$ thus represents the occupation of the eigenstate
$\ket{n}$ of the density matrix.
\begin{table}[!b]
\begin{center}
\renewcommand{\arraystretch}{1.2}
\begin{tabular*}{1.0\linewidth}
  {@{\extracolsep{\fill}}c|cccc} \hline\hline
                  & Formula & $L$    &  $S/S_{\text{max}}$  & $E_{\text{gap}}$ (eV)\\
\hline
thiophene         & C$_4$H$_4$S      &   5  & 0.07  & 11.19 \\ 
pyridine          & C$_5$H$_5$N      &   6  & 0.11  & 10.61 \\  
benzene           & C$_6$H$_6$       &   6  & 0.10  & 11.39 \\  
benzene (Hubbard) &      -           &   -  & 0.50  &  -    \\  
biphenyl          & C$_{12}$H$_{10}$ &  12  & 0.10  & 9.24  \\ 
naphthalene       & C$_{10}$H$_{8}$  &  10  & 0.11  & 8.65  \\ 
anthracene        & C$_{14}$H$_{10}$ &  14  & 0.12  & 7.06  \\ 
OPV2              & C$_{14}$H$_{12}$ &  14  & 0.10  & 8.30  \\ 
\hline\hline 
\end{tabular*}
\end{center}
\caption{\label{tab:Entropies} Chemical formula, number of $p_z$ orbitals ($L$)
  included in the PPP model and exact ground state entropies ($S$) for the listed
molecules.}
\end{table}

We note that $0\leq S\leq L\log{2}$, where $2L$ is the dimension of the
single-particle Hilbert space including spin. The expression for
$S_{\text{max}}$ follows because the number of electrons equal $L$ in all the
systems, i.e. half filled ``band''. When $\ket{\Psi_0}$ is a single Slater
determinant (corresponding to zero correlation) we have $S=0$, and when
$\ket{\Psi_0}$ has equal weight on a complete set of orthogonal Slater
determinants (corresponding to maximal correlation) we have $\rho_n = 1/2$ for
all $n$ and thus $S=L\log{2}$. Thus the number $0\leq S/S_{\text{max}}\leq 1$
represents a natural measure of the degree of correlation in $\ket{\Psi_0}$.

\section{Results}\label{sec:results}

\subsection{Total energies}\label{sec:total}
We first address the degree of correlation in the exact ground states by
considering the von Neumann entropies of the corresponding density matrices. The
calculated entropies are listed in Tab.~\ref{tab:Entropies}. Except for the
Hubbard description of benzene (see Sec. \ref{sec:hubbard}) which clearly
presents strong correlations, the entropies of the ground states are $\sim 10\%$
of their maximum value $S_{\text{max}}$ corresponding to weakly correlated
systems. The finite values of the entropies reveal that none of the ground
states are single Slater determinants implying that the Hartree-Fock ground
state energies will be larger than the exact ones.

\begin{figure}[!h]
\begin{center}
\includegraphics[width=0.9\linewidth]{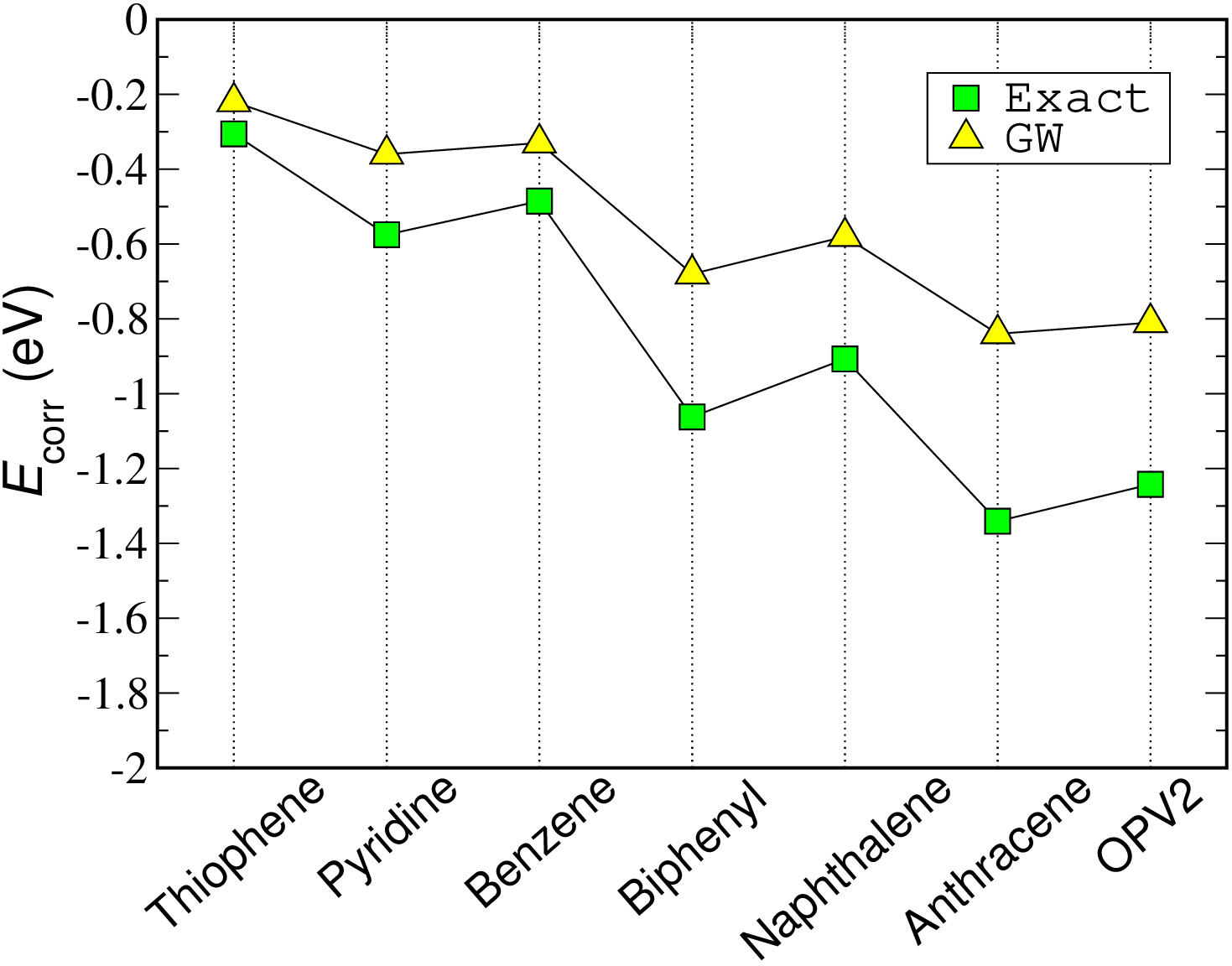}
\end{center}
\caption{\label{fig1} (Color online) Exact and GW correlation energies of the neutral
  groundstate of the seven molecules.}
\end{figure}

We here follow the usual convention and define the correlation energy as the
part of the total energy not included in Hartree-Fock, i.e.
\begin{equation}
  \label{eq:CorrelationEnergy}
  E_{\text{corr}} = E_{\text{exact}} - E_{\text{HF}} .
\end{equation}
Fig.~\ref{fig1} shows the exact correlation energies of the neutral molecules
together with those obtained by evaluating the total energy from Eqs.
\eqref{eq.E0} and \eqref{eq.xc} with the self-consistently determined Green
function and GW self-energy.

For the series of molecules considered here the correlation energy constitute
less than $0.5\%$ of the total energies. Furthermore, as expected it decreases
(in absolute size) with the number of atoms in the molecule. Clearly, the GW
approximation performs reasonably well for all the molecules capturing on
average $66\%$ of the correlation energy.

\subsection{Spectral properties}\label{sec:spec}

\begin{figure}[!h]
\begin{center}
\includegraphics[width=.9\linewidth]{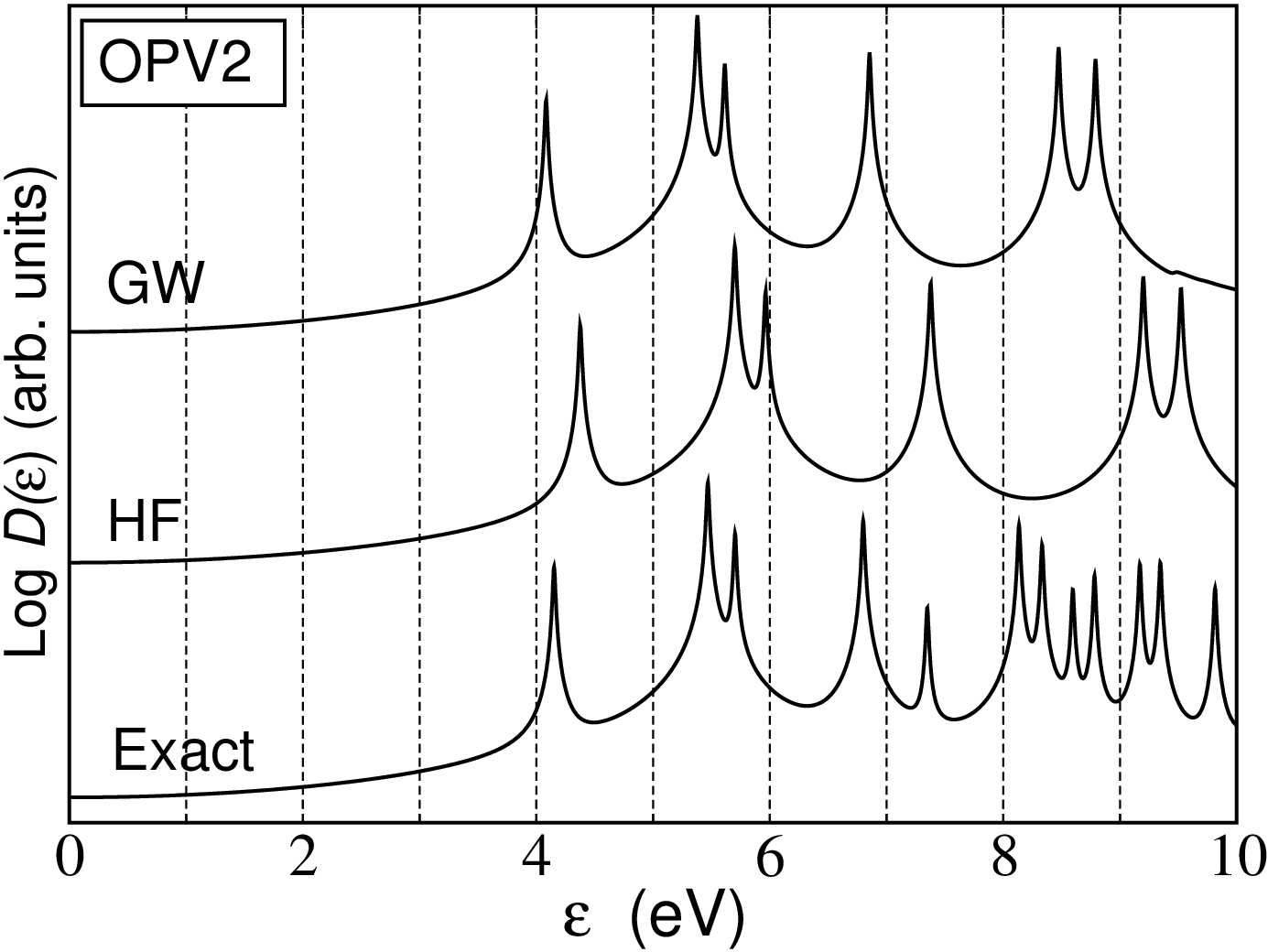}
\end{center}
\caption{\label{fig4} (Color online) Single-particle DOS of the OPV2 molecule.
  Note the logarithmic axis.}
\end{figure}

For isolated systems such as molecules, true quasi-particles resembling
single-particle excitations are characterized by having a weight close to unity
(for non-degenerate levels) in the spectral function, i.e.
\begin{equation}
  \label{eq:QuasiWeight}
  Z_n = \sum_i \vert \bra{\Psi_n^{N+1}}c_i^{\dagger}\ket{\Psi_0^N}\vert^2
        \sim 1 .
\end{equation}
This is equivalent to saying that there exists an orbital $|\nu\rangle$ so that
the excited state $\bra{\Psi_n^{N+1}}$ can be written as the single-particle
excitation $c^\dagger _{\nu}\ket{\Psi_0^N}$. In Fig.~\ref{fig4} we show the
single-particle density of states (DOS),
\begin{equation}
D(\varepsilon)=\sum_i A_i(\varepsilon)
\end{equation}
for the OPV2 molecule on a logarithmic scale. The height of the peaks reflects
the value of $Z_n$ (modulo degeneracies). The HF, and in particular, the GW
approximation reproduce the lowest lying excitations quite well while higher
excitations are poorly described. All the peaks in the HF spectrum have $Z_n=1$
while GW does shift some spectral weight from the main peaks to tiny satellite
structures (at higher energies than shown on the plot). However, the GW
satellites do not correspond to features in the exact spectrum. This shows that
excitations with $Z_n \ll 1$, i.e. excitations which do not have single-particle
character, are not well described by GW whose main effect is to improve the
position of the HF single-particle peaks.

In the following we consider the lowest lying single-particle excitations of the
molecules as obtained with Hartree-Fock, G$_0$W$_0$ and self-consistent GW. In
the G$_0$W$_0$ calculations the starting Green function G$_0$ is taken to be the
self-consistently determined Hartree-Fock Green function. Fig.~\ref{fig3} gives
an overview of the calculated excitation energies relative to the exact ones.
Energies corresponding to electron removal and electron addition are located on
the negative and positive half of the $x$-axis, respectively. From this plot
clear trends in the calculated excitation energies emerge.

Within HF the occupied (unoccupied) levels are systematically
overestimated (underestimated), and the deviation from the exact values worsens for
the higher lying excitations. A closer inspection of the figure reveals a few HF
energies at $\sim\pm$5.0 eV and $\sim\pm$5.7 eV that more or less coincide with
the exact energies. These are the HOMO and LUMO levels of the small single-ring
molecules thiophene, pyridine and benzene. The good agreement with the exact
levels for these systems does \emph{not} arise because HF gives a correct description of
the many-body states and their energies. This was already clear from the
analysis above which showed that the exact eigenstates are not single Slater
determinants and hence the excitation energies in Eq.~\eqref{eq:QuasiLevels}
have contributions from both $\Delta_{\text{relax}}$ and $\Delta_{\text{corr}}$.
The good agreement must therefore be ascribed to cancellations between the
relaxation and correlation contribution to the exact energies (this is discussed
further in connection with Fig. \ref{fig3}).

Both the G$_0$W$_0$ and the GW give consistently better energies than HF -- in
particular for the higher lying excitations where the absolute errors are
reduced to less than $\sim$0.4 eV as compared to $\sim$1 eV for HF. For the
low-lying excitations GW slightly overestimates (underestimates) the occupied
(unoccupied) levels corresponding to an overcorrection of the HF energies.

\begin{figure}[!h]
\begin{center}
\includegraphics[width=.95\linewidth]{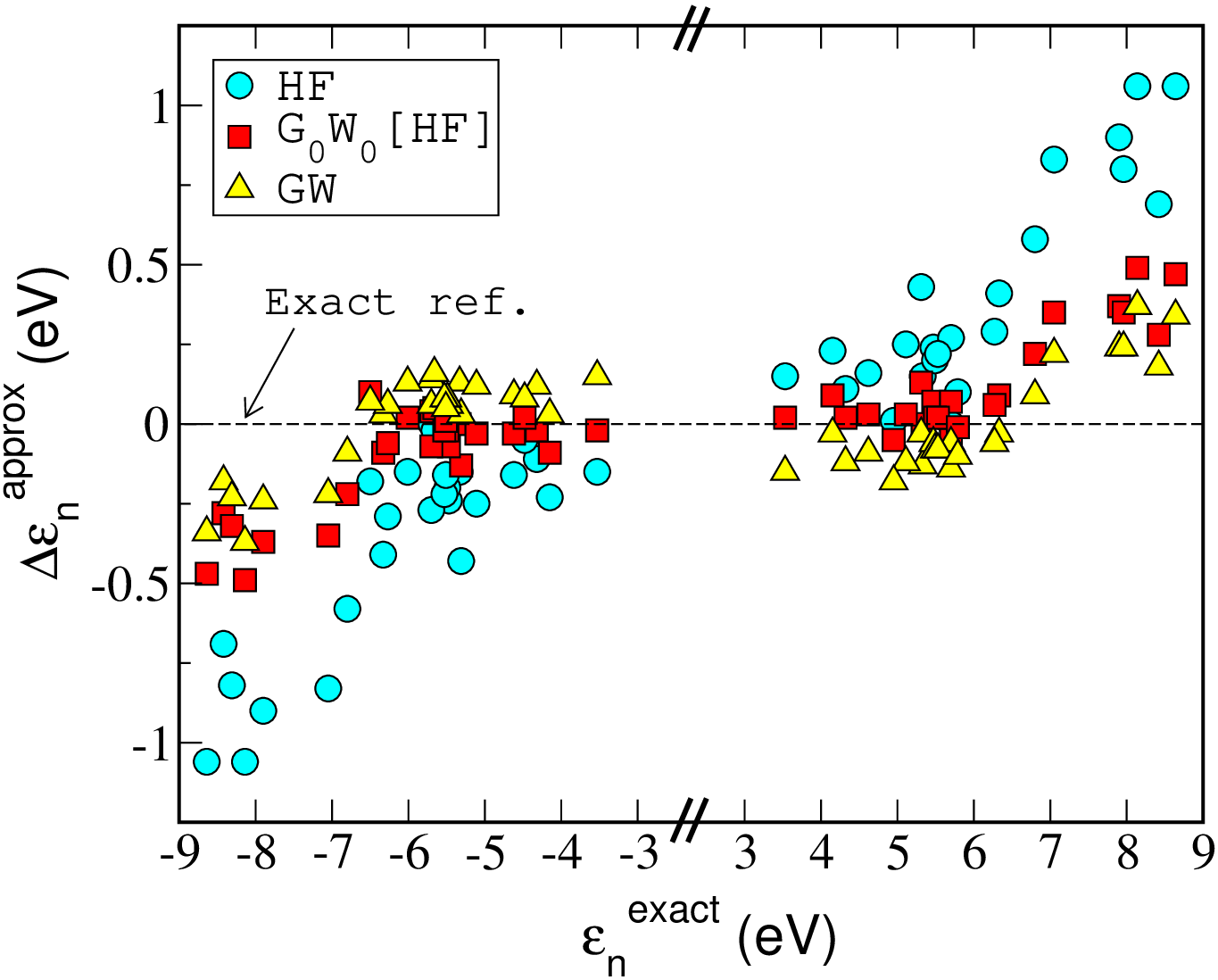}
\end{center}
\caption{\label{fig2} (Color online) Energy of the 3 highest occupied and 3
  lowest unoccupied molecular orbitals relative to the exact values. While
  Hartree-Fock underestimates the occupied and overestimates the unoccupied
  levels, self-consistent GW shows the opposite trends but deviates on average
  less from the exact result.}
\end{figure}

In order to address the relative contributions from $\Delta_{\text{relax}}$ and
$\Delta_{\text{corr}}$ to the excitation energies in Eq.~\eqref{eq:QuasiLevels},
we plot in Fig.~\ref{fig3} the difference between the exact gaps and the gaps
obtained from the (i) Hartree-Fock eigenvalues, (ii) Hartree-Fock total energy
differences \emph{with} self-consistent relaxations in the $N\pm 1$ Slater
determinants taken into account, and (iii) the distance between the highest
occupied and lowest unoccupied peaks in the GW spectral function. By using the
expression for the quasi-particle energies in Eq.~\eqref{eq:QuasiLevels}, the
exact gap $E_{\text{gap}} = \varepsilon_{\text{LUMO}} -
\varepsilon_{\text{HOMO}}$ can be expressed as
\begin{equation}
  \label{eq:Gap}
  E_{\text{gap}} = \varepsilon_{\text{LUMO}}^{\text{HF}} - 
                   \varepsilon_{\text{HOMO}}^{\text{HF}} + 
                   \Delta_{\text{relax}}^{\text{gap}} + 
                   \Delta_{\text{corr}}^{\text{gap}}
\end{equation}
where $\Delta_{\text{relax}}^{\text{gap}}$ and
$\Delta_{\text{corr}}^{\text{gap}}$ are the gap equivalents of the corresponding
quantities in Eq.~\eqref{eq:QuasiLevels} and
$\varepsilon_{\text{HOMO/LUMO}}^{\text{HF}}$ are the Hartree-Fock HOMO/LUMO
eigenvalues. By definition $\Delta_{\text{relax}}^{\text{gap}}$ is difference
between the gaps obtained from the HF eigenvalues and relaxed HF total energy
differences. In Fig.~\ref{fig3} this is given by the vertical distance between
the (blue) squares and circles. The correlation contribution
$\Delta_{\text{corr}}^{\text{gap}}$ can be read off as the difference between
the exact gap (dashed horizontal line) and the relaxed HF total energy gap (blue
squares). Inclusion of relaxation effects clearly reduces the HF gaps
considerably implying that $\Delta_{\text{relax}}^{\text{gap}}<0$. This
reduction is due to the screening from the orbital relaxation which reduces the
Coulomb interaction with the added hole or electron and hence also the gap.


\begin{figure}[!h]
\begin{center}
\includegraphics[width=.9\linewidth]{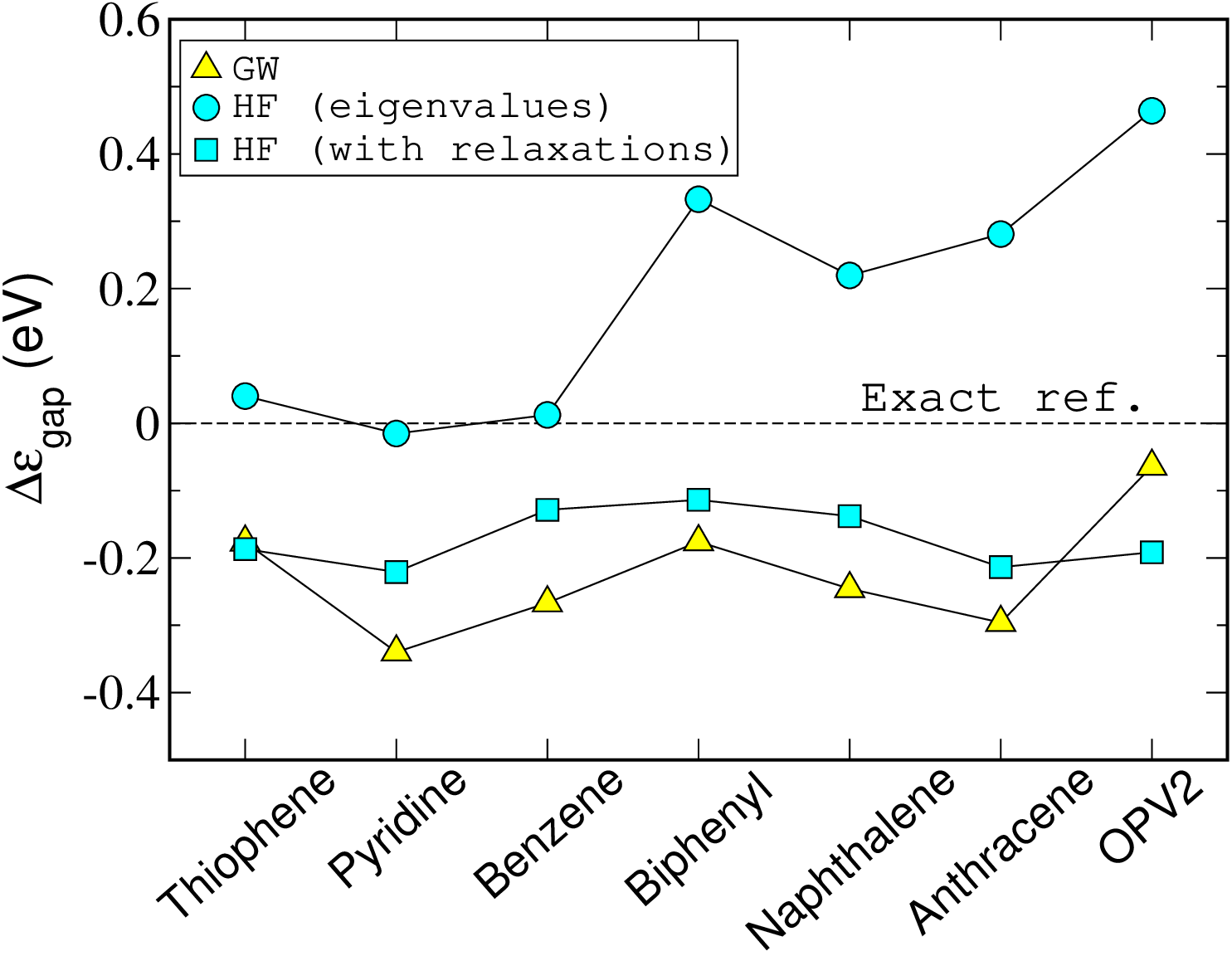}
\end{center}
\caption{\label{fig3} (Color online) The HOMO-LUMO gap relative to the exact
  values. In addition to the HF and GW single-particle energies, the relaxed
  Hartree-Fock total energy differences,
  $E_0^{\text{HF}}(N+1)+E_0^{\text{HF}}(N-1)-2E_0^{\text{HF}}(N)$ are also
  shown. The excellent results of HF for the three smallest molecules is a
  result of error cancellation between relaxation and correlation
  contributions.}
\end{figure}

In contrast to the HF (eigenvalue) gaps for which the agreement with the exact
gap worsens as a function of the size of the molecules, the GW gaps follow more
consistently the same trend and underestimates the exact gaps with $0.05-0.35$
eV for all the molecules. The close resemblance between GW and the relaxed HF
result indicates that the effect of GW is mainly to account for the screening
effects included in HF via orbital relaxations, $\Delta_{\text{relax}}$.

\subsection{Long- versus short-range interactions}\label{sec:hubbard}

To demonstrate the shortcomings of the GW approximation for strongly correlated
systems, we consider a Hubbard model description of the benzene molecule. It
should be noted that this Hubbard description of benzene is not intended as a
realistic description of the benzene molecule, rather it serves to illustrate
the limitations of the GW approximation. The Hamiltonian is identical to the
PPP-Hamiltonian in Eq.~\eqref{eq:PPPHamiltonian}, except that the long range
Coulomb interactions in the third term have been omitted. The values for the
hopping elements and the onsite Coulomb interaction are $t=2.539$ and $U=10.06$,
respectively. With a $U/t$-ratio of $\sim$4 this obviously represent a strongly
correlated system. The latter is reflected in the ground state entropy in
Tab.~\ref{tab:Entropies} which is $50\%$ of its maximum value.

From the calculated total energies we find that the correlation energy (not
included in Fig.~\ref{fig1}) constitutes $10\%$ of the ground state energy which
is a considerably higher fraction as for the PPP descriptions of the molecules.
The GW total energy captures $88\%$ of the correlation energy compared to $66\%$
on the average for the PPP descriptions. However, from an absolute point of
view, the GW approximation misses the exact ground state energy by $0.48$ eV.
This should be compared to $0.16$ eV which is the difference between the exact
and the GW ground state energy for the PPP description of benzene.

\begin{figure}[!h]
\begin{center}
\includegraphics[width=.9\linewidth]{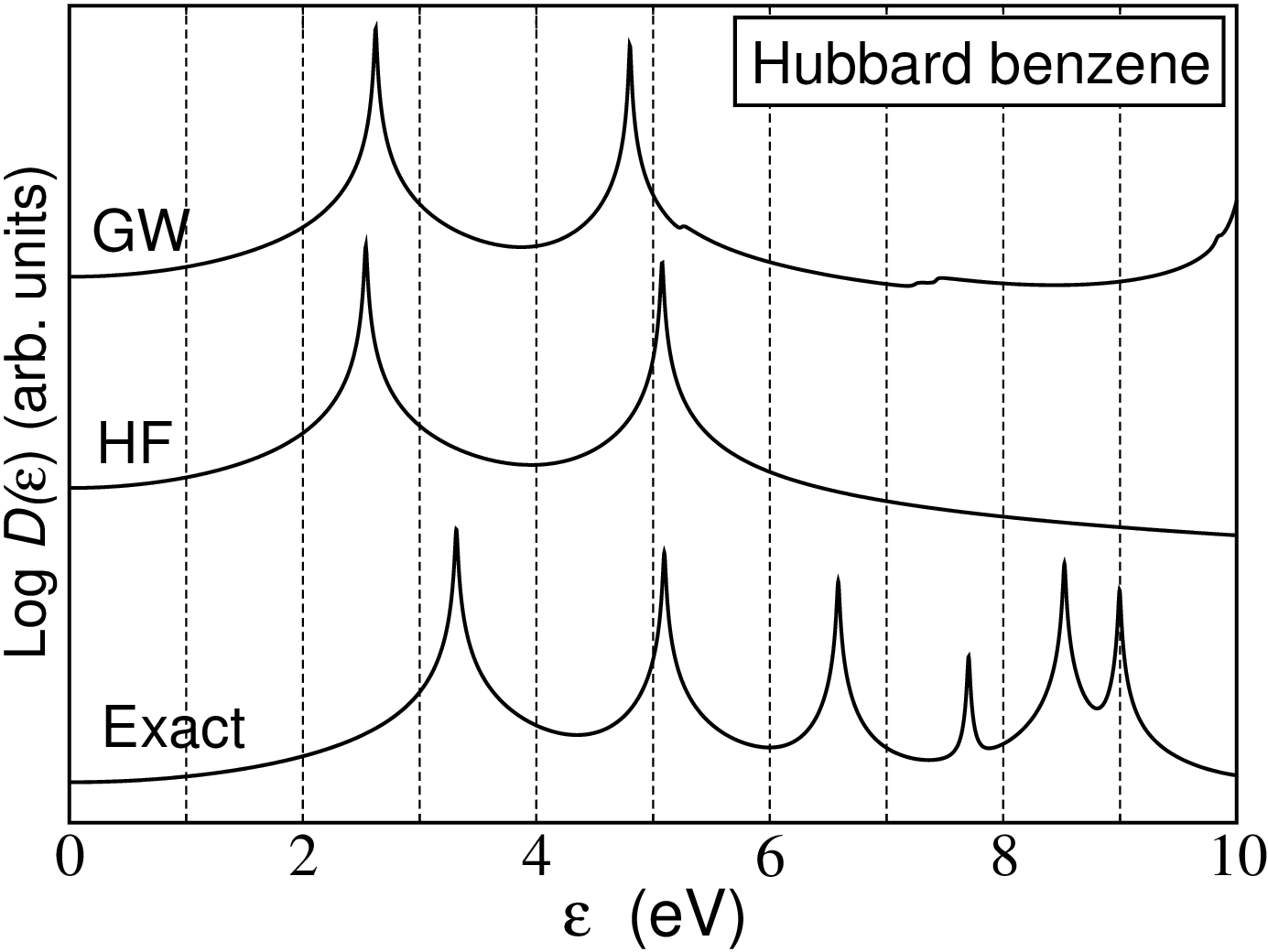}
\end{center}
\caption{\label{fig5} (Color online) Single-particle DOS for the Hubbard
  description of the benzene molecule (only on-site interactions from the PPP
  model are kept). Note the logarithmic axis.}
\end{figure}

The poor performance of both Hartree-Fock and GW for the spectral properties of
the Hubbard benzene is illustrated in Fig.~\ref{fig5} which shows the spectral
function as calculated with the two methods together with the exact one. Both
Hartree-Fock and GW severely underestimates the position of the LUMO level and
completely misses the details of the spectrum at higher energies.

This clearly demonstrates that GW is of limited relevance when considering
systems where correlation effects ($\Delta_{\text{corr}}$) dominates over
screening, or relaxation, effects ($\Delta_{\text{relax}}$).

\subsection{Exact Kohn-Sham orbital energies}\label{sec:latticedft}

Within density functional theory (DFT) the eigenvalues of the
single-particle Kohn-Sham Hamiltonian, $H_{\text{KS}}$, are often interpreted as physical energies. In principle the
validity of this approximation depends on the size of the derivative
discontinuity of the true and unknown exchange-correlation (xc-)
functional\cite{perdew_levy,godby_sham}. In practice the use of
semi-local xc-functionals represents an additional
approximation. It is of general interest to investigate to what extent the disrepancy between
KS energies and true QP energies result from the use of approximate
functionals and to what extent this is a property of the exact functional.
Below we compare the \emph{exact} Kohn-Sham spectrum to the exact QP spectrum of the
PPP benzene molecule using the lattice version of DFT.

The lattice version of DFT follows by extending the fundamental concepts of
standard DFT, such as the Hohenberg-Kohn theorem and the Kohn-Sham equations, to
model Hamiltonians as e.g. the PPP-Hamiltonian.~\cite{Gunnarsson-LatticeDFT} In this reformulation of DFT the
site occupations $n_i$ replaces the continuous electron density $n(\mathbf{r})$
as the fundamental variable that determines the ground state properties. The
lattice version of the single-particle Kohn-Sham Hamiltonian is
given by the sum of the hopping terms (the kinetic energy) and a site dependent
Kohn-Sham potential $V_i^{\text{KS}}$ which is constructed to yield the correct
site occupations of the ground state,
\begin{equation}
  \label{eq:H_KS_lattice}
  H_{\text{KS}} = - \sum_{\langle ij\rangle\sigma}t_{ij} 
                    c^{\dagger}_{i\sigma}c^{\phantom\dagger}_{j\sigma}
                  + \sum_i V_i^{\text{KS}} \hat{n}_i .
\end{equation}
For the present purpose the explicit form of the site potential
$V_i^{\text{KS}}$ is not important. The fact that the lattice version of the
Kohn-Sham potential is an onsite potential, is equivalent to the restriction of
the Kohn-Sham potential $V_{\text{xc}}(\mathbf{r})$ in the real-space
formulation of DFT to a local potential.

Due to the high symmetry of the benzene molecule all sites in the
PPP-Hamiltonian are equivalent implying that $V_i^{\text{KS}}$ has the
same value for all sites. Except for a constant shift, the eigenvalues
of the Kohn-Sham Hamiltonian are therefore given by those of the
hopping part of the Hamiltonian. The HOMO-LUMO gap calculated from the
Kohn-Sham eigenvalues is $E_{\text{gap}}^{\text{KS}}=5.08$~eV which is
a severe underestimation of the true gap of $11.39$~eV. In line with previous
studies\cite{perdew_levy,godby_sham} we thus conclude that the main
reason for the discrepancy between KS eigenvalues obtained with approximate xc-functionals and the exact orbital energies is
due to the derivative discontinuity of the exact xc-functional.

\section{Conclusion}\label{sec:conclusions}
We have presented calculations for the total energy and charged
single-particle excitations in seven conjugated molecules described by
the semi-empirical PPP model within fully self-consistent GW and exact
diagonalization. The results show that the GW approximation gives a
consistently good description of both total energies and electronic
excitations with a slight tendency to overestimate (underestimate) the
position of the latter for occupied (unoccupied) levels. We have found
that the effect of the GW self-energy is similar to the inclusion of
orbital relaxations in the $N\pm 1$ final states in Hartree-Fock
theory. On the other hand the contribution to the excitation energies
coming from correlations in the ground- and excited states is less
well described by GW. This explains why GW tend to reduce electron
addition/removal energies relative to the HF eigenvalues. It was shown
that GW does not perform well for systems with short range
interactions (Hubbard models) where correlation effects are dominating
over screening/relaxation effects. Finally it was shown that the exact
Kohn-Sham eigenvalues significantly underestimate the true HOMO-LUMO
gap of a benzene molecule showing the importance of the derivative
discontinuity of the exact exchange-correlation fucntional.

\section{Acknowledgements}
Financial support from the Danish Council for Production and Technology (FTP)
under Grant 26-04-0181 ``Atomic scale modeling of emerging electronic devices''
is acknowledged. KST acknowledges support from the Danish Center for Scientific
Computing. The Center for Atomic-scale Materials Design (CAMD) is sponsored by
the Lundbeck Foundation.

\bibliographystyle{apsrev}   

\end{document}